\documentclass[intlimits,twoside,a4paper]{article}

\usepackage[cp1251]{inputenc}
\usepackage[eqsecnum]{cmpj3}
\usepackage{bm}

\usepackage{color, soul}

\issue{2021}{24}{2}{23705}
\doinumber{10.5488/CMP.24.23705}

\title[Mn ions' site and valence in PbTiO$_{3}$]%
{Mn ions' site and valence in PbTiO$_{3}$ based on the native
	vacancy defects%
	}
\author[H. Xin \textsl{et al.}]{H. Xin\orcid{0000-0002-1021-2221}\thanks{
	Email address: xinhong@xauat.edu.cn.}, Q. Pang, D.~L. Gao, L. Li, P. Zhang, J. Zhao
}
\address{
	 College of Science, Xi’an University of Architecture and Technology, Xi’an 710055, China
}

\Keywords{Mn doping, PT, native defects, defect formation energy, Bader charge}

\date{Received January 28, 2021, in final form April 15, 2021}
\begin{document}
	
	\maketitle

\begin{abstract}

Mn ions' doping site and valence were studied in PbTiO$_{3}$ (PT) with
the native vacancy defects by the first-principles calculations. Firstly, the
native vacancy defects of Pb, O and Ti in PT were investigated and it was found
that Pb vacancy is preferred to others. And then the growth of Mn doped PT should be
preferred to Mn ion substituting for an A-site Pb ion with +3 valence when Pb is
deficient under equilibrium conditions driven solely by minimization of the
formation energy, and this could result in a larger lattice distortion of PT. In
addition, when Mn enters the Pb site, the electronegativity of O becomes weaker
which makes the domain movement easier in PT to improve the performance of PT,
while Mn ion substitution for a B-site Ti ion is the opposite.
\printkeywords
%
\end{abstract}

\section{Introduction}

The advantage of lead titanate PbTiO$_{3}$ (PT) and lead zirconate titanate
PbZr$_{1-x}$Ti$_{x}$O$_{3}$ (PZT) as Pb-based ferroelectric materials, has been
well exhibited in infrared pyroelectric sensors, piezoelectric transducers,
nonvolatile memories and so on~\cite{1,2,3,4}. Furthermore, some cations were also
introduced to modify their properties and improve their applications~\cite{5,6,7,8,9,10,11,12,13,14,15}. Mn
doped PT and PZT exhibiting different characteristics with dopant concentration have
attracted a lot of interest~\cite{14,15,16,17,18}. Especially, Mn can have a softening effect
on PT and PZT at a small concentration~\cite{14,16,17}. It was deemed that the
occurrence of a small amount of Mn ions entering the A site was compatible with
small differences between the calculated and measured x-ray absorption
near-edge structures (XANES)~\cite{19}. Then, it acts as a B-site dopant to
decrease the relative permittivity and loss tangent with the concentration
increasing. The effect of Mn doping on the structure of PT and PZT is still not
well understood.

Mn defects could always occur with native defects simultaneously during the
formation process of Mn doped PT and PZT. Several native defects have been reported, such
as Pb vacancy~(V$_{{\rm{Pb}}}$)~\cite{20}, O~vacancy~(V$_{{\rm{O}}}$)~\cite{20} and Ti vacancy~(V$_{{\rm{Ti}}}$)~\cite{21}, which may affect the role of Mn in PT and PZT. Thus, the study of Mn doped PT and PZT
should be carried out based on the native defects. In this paper Mn defect in PT is studied 
by the first-principles calculation due to their similar structure,
which would be helpful to clarify the influence of Mn doping on PZT.

In this paper, the native vacancy defects of PT were investigated by the
first-principles calculation firstly. Their stability was analyzed through the
defect formation energy which is a function of Fermi-energy under different
growth conditions. Then, Mn defect position and valence state in PT were
determined based on the most probable native vacancy defects. Additionally, the
electronegativity of O was investigated in PT without and with Mn doping at
different sites, and its effect on the performance of PT was discussed too.

\section{Model structures and computational methods}

\subsection{Model structures}

\begin{figure}[h]
	\centering
	\includegraphics[width=0.65\linewidth]{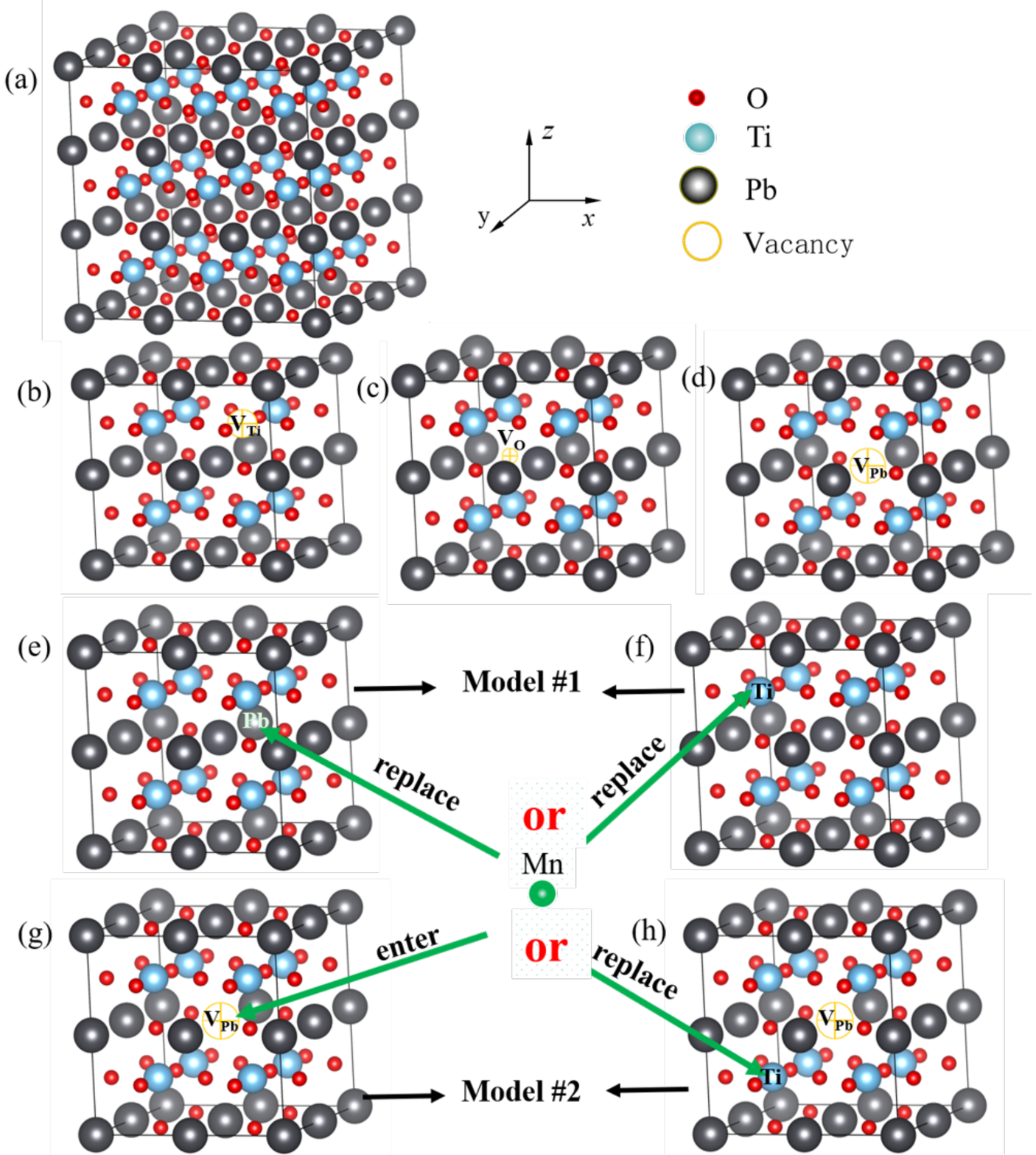}
	\caption{(Colour online) The model structures PT without and with different
		defects, (a) ${\rm{3}} \times {\rm{3}} \times {\rm{3}}$ perfect supercell, (b) to
		(h) ${\rm{2}} \times {\rm{2}} \times {\rm{2}}$ supercells for different defects,
		(b) V$_{\rm{Ti}}$, (c) V$_{\rm{O}}$, (d) V$_{\rm{Pb}}$, and (e) to (h) schematics showing Mn
		doping into Pb site and Ti site in two kinds of host supercells, respectively, (e)
		and (f) perfect supercells marked with model \#1, (g) and (h) supercells with one
		V$_{\rm{Pb}}$ marked with model \#2. Red balls, blue balls, grey balls, hollow yellow
		balls and a green ball denote O, Ti, Pb ions, Pb vacancy
		(${{\rm{V}}_{\rm{Pb}}}$) and a Mn ion, respectively.}
	\label{fig-smp1}
\end{figure}

The cubic phase of PT (space group Pm3m), which is often used in the first principles simulations, was chosen as a model to simplify the calculation~\cite{22}. For
simulations, a ${3} \times {3} \times {3}$ supercell [shown in
figure~\ref{fig-smp1}~(a)] composed of 135 atoms and 27 primitive PT unit cells was established
by repeating the unit cell. Then, one vacancy defect (such as V$_{\rm{Pb}}$,
V$_{\rm{O}}$ and V$_{\rm{Ti}}$) was introduced into the ${\rm{3}} \times {\rm{3}} \times
{\rm{3}}$ supercell for the calculation of the native vacancy defect. It is difficult to show the structure of defects in PT with the dense
ion arrangement. Thus, $2 \times
2 \times 2$ supercells were used to illustrate the structure of defects in PT in
figures~\ref{fig-smp1}~(b)~to~(h). Additionally, the model structures of Mn defects in PT were
built by adding one Mn ion into two kinds of host supercells, one ${\rm{3}}
\times {\rm{3}} \times {\rm{3}}$ perfect supercell of PT [marked with model \#1 in figures~\ref{fig-smp1}~(e)
and~\ref{fig-smp1}~(f)] and the other containing one vacancy defect [marked with model \#2 in
figures~\ref{fig-smp1}~(g) and ~\ref{fig-smp1} (h) where V$_{\rm{Pb}}$ was taken as an example]. The
concentrations of native defects and Mn defects are all 3.7\% in the calculation,
while it is about 3\% for Mn doping in the experiment~\cite{16}.

\subsection{Computational methods}

\subsubsection{First-principles}

All the total energy calculations were performed in VASP code using the
projector-augmented wave (PAW) method~\cite{23,24}. Generalized gradient
approximations (GGA) were used in the form of Perdew-Burke-Ernzerhof (PBE)
exchange-correlation function for the electron-electron interaction~\cite{25}. A
Monkhorst-Pack mesh and a plane-wave cutoff energy were determined at $6 \times
6 \times 6$ and 500~eV, respectively, through calculating the total energy. The
convergence of the electronic self-consistent energy less than $10^{-5}$~eV and
the force on each atom less than 0.01~eV/\AA{} were defined for full relaxation of
the positions of all atoms.

The optimized lattice constant of PT is 3.972~\AA{}, which is very close to
the experimental value of~3.97~\AA{}~\cite{22}. The band gap of PT was calculated to be
2.40~eV in this paper, which is lower than the experimental value of~3.40~eV~\cite{26}. The band gap is always undervalued with the first-principles~\cite{27,28}, so
the experimental value of~3.40~eV was used in calculation presented below.

\subsubsection{Defect formation energies}

The defect formation energy ${{E}_f}({D^q})$ of a defect $D$ in the
charge state $q$ is defined as~\cite{29,30}:

\begin{align}
{E_f}({D^q}) = {E_\text{tot}}({D^q}) - {E_\text{tot}} \pm \sum\limits_i {n{}_i} {{\rm{\mu
	}}_i} + q({{\rm{\varepsilon }}_\text{F}} + E{}_V + \Delta V),
\label{eq:1}
\end{align}
where ${E_\text{tot}}({{D}^q})$ and ${E_\text{tot}}$ are the total energy of the
supercell with and without defect, respectively, ${n_i}$~is the number of $i$ atoms
which are removed from (or added to) the supercell corresponding to plus
(or minus) sign, ${\mu _i}$ is its chemical potential, $q$ is the charge state of the
defect, ${\varepsilon _\text{F}}$ is Fermi-energy which is calibrated to 0 at the top of the valence band ${E_V}$. Finally, a correction term $\Delta V$ is added to align the reference potential in the
defect supercell with that in the bulk.

When the defect formation energies ${E_f}({D^{{q_1}}})$ and ${E_f}({D^{{q_2}}})$ in two charge states of ${q_1}$ and ${q_2}$ are equal,
the defect transition Fermi-energy ${\varepsilon _{{q_1} - {q_2}}}$, which is defined as the thermodynamic
transition level~\cite{29}, can be determined as the following form according to
equation~(\ref{eq:1}):

\begin{align}
{\varepsilon _{{q_1} - {q_2}}} = \frac{{{E_\text{tot}}({D^{{q_1}}}) - {E_\text{tot}}({D^{{q_2}}})}}{{{q_2} - {q_1}}} - {E_V} - \Delta V.
\label{eq:2}
\end{align}

To grow the crystal of PbTiO$_{3}$ in thermodynamic equilibrium conditions, ${\mu _{\rm{Pb}}}$, ${\mu _{{\rm{Ti}}}}$, 
${\mu _{\rm{O}}}$ and ${\mu _{{\rm{PbTi}}{{\rm{O}}_{\rm{3}}}}}$ (the chemical potentials of Pb, Ti, O and PbTiO$_{3}$) must be satisfied as follows:
\begin{align}
{\mu _{\rm{Pb}}} + {\mu _{{\rm{Ti}}}} + 3{\mu _{{\rm{O}}}} = {\mu _{{\rm{PbTi}}{{\rm{O}}_3}}}.
\label{eq:3}
\end{align}
PbO$_{2}$, PbO and TiO$_{2}$ are considered to be major competing phases. The
following conditions need to be further taken into account in order to inhibit
the undesired phases in PbTiO$_{3}$~\cite{31}
\begin{align}
{\mu _{{\rm{Pb}}}} &\leqslant 0,\;\;{\mu _{{\rm{Ti}}}} \leqslant 0,\;\;{\mu _{\rm{O}}} \leqslant 0,\; \notag \\
{\mu _{{\rm{Pb}}}} &+ 2{\mu _{\rm{O}}} < {\mu_{{\rm{Pb}}{{\rm{O}}_{\rm{2}}}}},\notag \\
{\mu _{{\rm{Pb}}}} &+ {\mu _{\rm{O}}} < {\mu _{{\rm{PbO}}}},\notag \\
{\mu _{{\rm{Ti}}}} &+ 2{\mu _{\rm{O}}} < {\mu_{{\rm{Ti}}{{\rm{O}}_{\rm{2}}}}},\notag \\
{\mu _{{\rm{Pb}}}} &+ {\mu _{{\rm{Ti}}}} \geqslant {\mu_{{\rm{PbTi}}{{\rm{O}}_{\rm{3}}}}},\;
\label{eq:4}
\end{align}
where ${\mu _{{\rm{Pb}}{{\rm{O}}_{\rm{2}}}}}$, ${\mu _{{\rm{PbO}}}}$ and ${\mu _{{\rm{Ti}}{{\rm{O}}_2}}}$ are the chemical potentials of corresponding competing phases.
Their corresponding bulk energies of per formula unit are listed in
table~\ref{tbl-smp1}.
\begin{table}[!b]
	\caption{Bulk energies of per formula unit of competing phases of PT.}
	\label{tbl-smp1}
\begin{center}
	\vspace{3pt} \noindent
	\begin{tabular}{p{104pt}p{104pt}}
		\hline
		\parbox{104pt}{\centering 
			{\small System}
		} & \parbox{104pt}{\centering 
			{\small ${E_\text{bulk}}$ (eV)}
		} \\
		\hline
		\parbox{104pt}{\centering 
			{\small PbTiO$_{3}$}
		} & \parbox{104pt}{\centering 
			{\small $-13.00$}
		} \\
		\parbox{104pt}{\centering 
			{\small TiO$_{2}$}
		} & \parbox{104pt}{\centering 
			{\small $-9.95$}
		} \\
		\parbox{104pt}{\centering 
			{\small PbO}
		} & \parbox{104pt}{\centering 
			{\small $-2.85$}
		} \\
		\parbox{104pt}{\centering 
			{\small PbO$_{2}$}
		} & \parbox{104pt}{\centering 
			{\small $-3.66$}
		} \\
		\hline
	\end{tabular}
\end{center}
\end{table}
\begin{figure}[!t]
	\centering
	\includegraphics[width=0.65\linewidth]{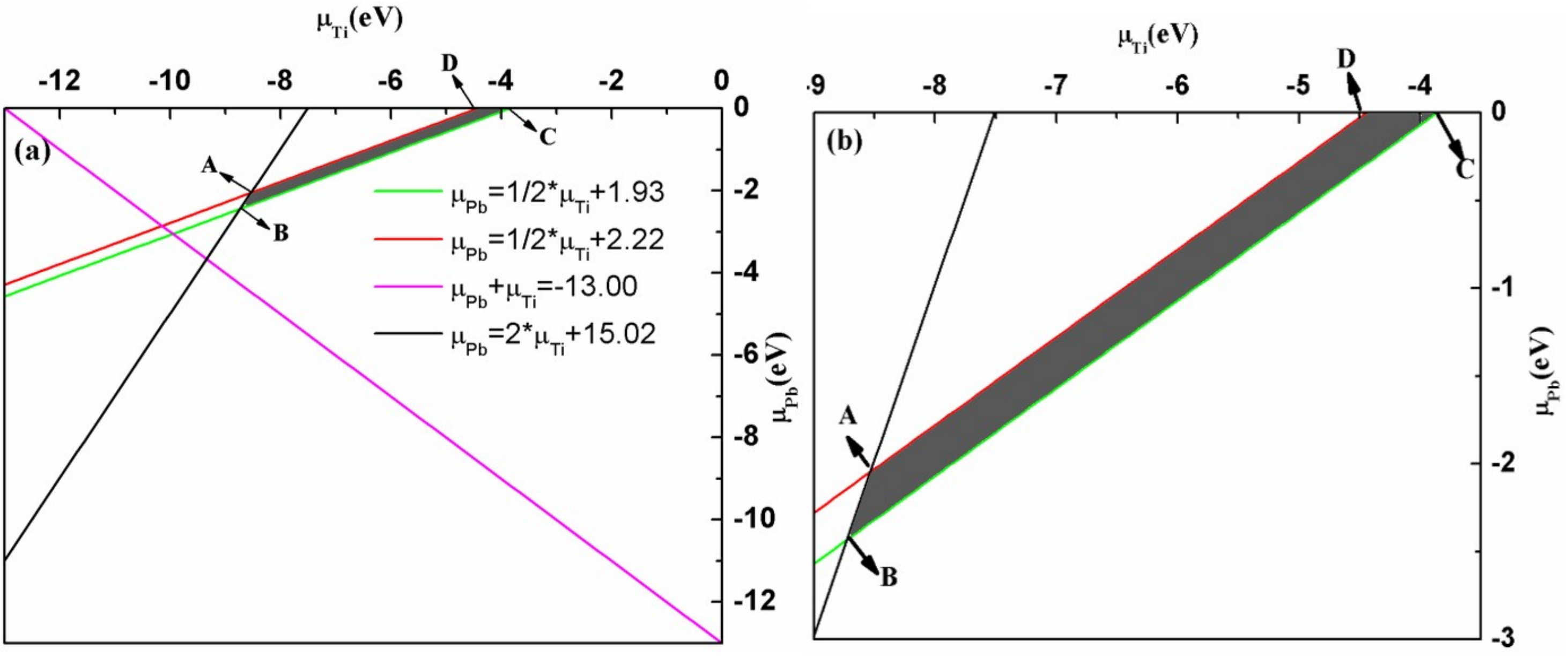}
	\caption{(Colour online) Region of stability (shaded) for PbTiO$_{3}$ in the
	2D space spanned by  ${\mu _{{\rm{Ti}}}}$ and ${\mu _{{\rm{Pb}}}}$, (a) full scale
	and (b) the magnified stable region. The (colored) lines are the limits from the
	compound of interest and the competing phases. A, B, C, and D are four vertices
	of that region corresponding to four different growth conditions of stable PT
	growth.}
	\label{fig-smp2}
\end{figure}

The chemical potentials of Pb, Ti and O, ${\mu _{{\rm{Pb}}}}$, ${\mu
	_{{\rm{Ti}}}}$ and ${\mu _{\rm{O}}}$, could be determined from equations~(\ref{eq:3}) and~(\ref{eq:4})
 for stable PbTiO$_{3}$ growth, which were shown on the ${\mu_{{\rm{Pb}}}}$-${\mu _{{\rm{Ti}}}}$ plane in figure~\ref{fig-smp2} with a shaded region.
Then, the chemical potentials at four vertices (labeled with A, B, C, and D in
figure~\ref{fig-smp2}, respectively) of the quadrilateral shadow region, listed in table~\ref{tbl-smp2},
were selected to analyze the defect formation energies, which correspond to four
different growth conditions of stable PT growth. A and B points represent
the oxygen-rich growth of PT, where ${\mu _{\rm{O}}}$ is higher, while C and D points
are oxygen-poor, which correspond to  ${\mu _{\rm{O}}}$ in table~\ref{tbl-smp2}. However, they are
just opposite for Ti and Pb. Since the amount of Mn doping was very small, the
change of chemical potential of Mn was neglected and the energy $-5.16$~eV of Mn
atom in their bulk crystal phase was used in the calculation. For simplicity, ${\rm{Mn}}_{{\rm{Pb}}}^{{x}+}$ and ${\rm{Mn}}_{{\rm{Ti}}}^{{x}+}$ were used to represent the defects of Mn ions in $+x$ valence
state entering the Pb site and Ti site, respectively, in this paper.

\begin{table}[!t]
	\caption{Chemical potentials of Ti, Pb and O (${\mu _{{\rm{Ti}}}}$, ${\mu _{{\rm{Pb}}}}$ and ${\mu _{{\rm{O}}}}$) at A, B, C and D four vertices of the shadow region in figure~\ref{fig-smp2}.}
	\label{tbl-smp2} 
\begin{center}	
		\vspace{3mm}
	\begin{tabular}{p{72pt}p{72pt}p{72pt}p{72pt}}
		\hline
		\parbox{72pt}{\centering 
			{\small Chemical potential}
		} & \parbox{72pt}{\centering 
			{\small ${\mu _{\rm{Ti}}}$ (eV)}
		} & \parbox{72pt}{\centering 
			{\small ${\mu _{{\rm{Pb}}}}$ (eV)}
		} & \parbox{72pt}{\centering 
			{\small ${\mu _{\rm{O}}}$ (eV)}
		} \\
		\hline
		\parbox{72pt}{\centering 
			{\small A}
		} & \parbox{72pt}{\centering 
			{\small $-8.56$}
		} & \parbox{72pt}{\centering 
			{\small $-2.05$}
		} & \parbox{72pt}{\centering 
			{\small $-0.80$}
		} \\
		\parbox{72pt}{\centering 
			{\small B}
		} & \parbox{72pt}{\centering 
			{\small $-8.73$}
		} & \parbox{72pt}{\centering 
			{\small $-2.43$}
		} & \parbox{72pt}{\centering 
			{\small $-0.61$}
		} \\
		\parbox{72pt}{\centering 
			{\small C}
		} & \parbox{72pt}{\centering 
			{\small $-3.86$}
		} & \parbox{72pt}{\centering 
			{\small 0}
		} & \parbox{72pt}{\centering 
			{\small $-3.047$}
		} \\
		\parbox{72pt}{\centering 
			{\small D}
		} & \parbox{72pt}{\centering 
			{\small $-4.44$}
		} & \parbox{72pt}{\centering 
			{\small 0}
		} & \parbox{72pt}{\centering 
			{\small $-2.85$}
		} \\
		\hline
	\end{tabular}
\end{center}
\end{table}

\section{Results and discussion}

\subsection{Analysis of native vacancy defects}

\begin{figure}
	\centering
	\includegraphics[width=0.7\linewidth]{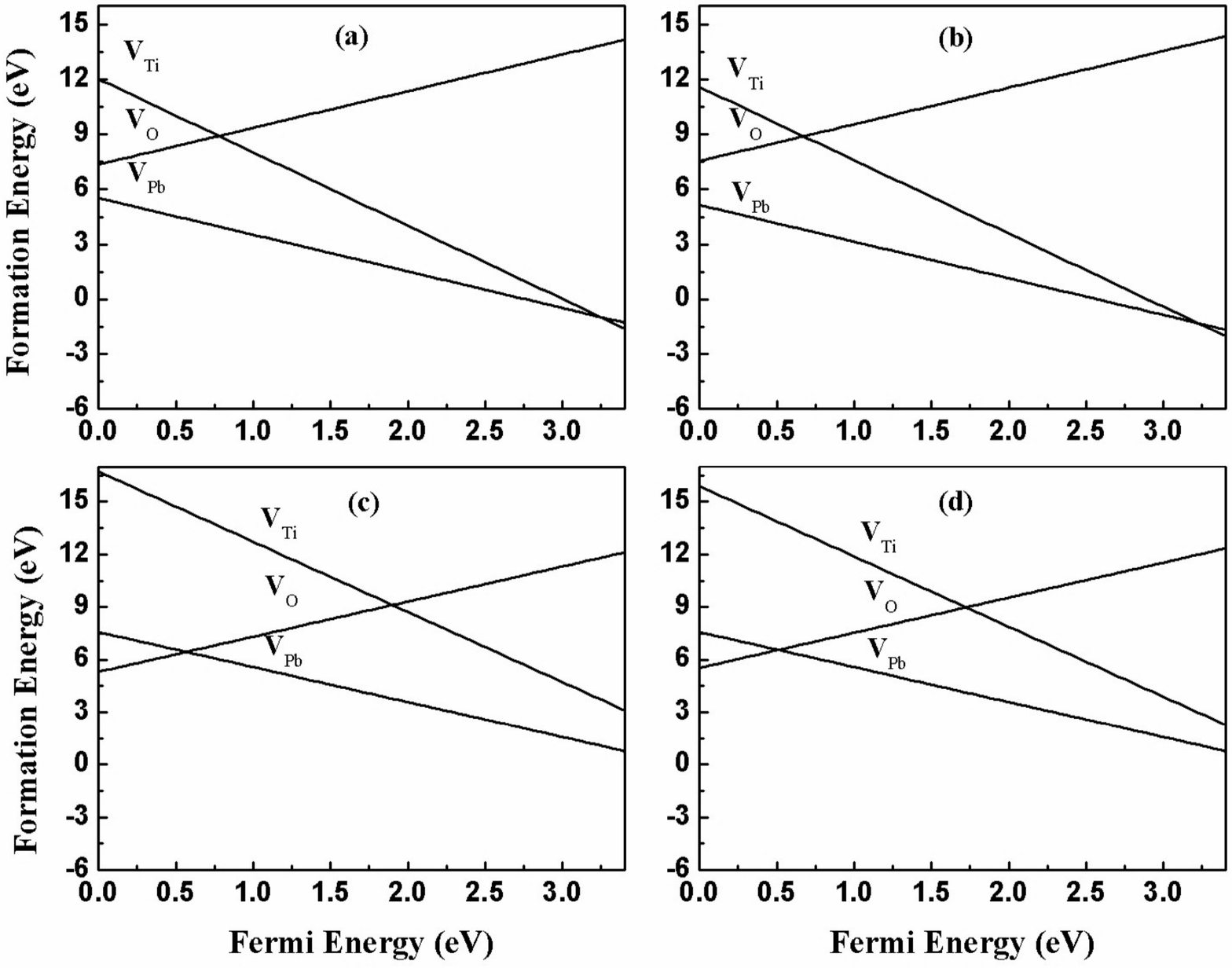}
	\caption{Formation energies as a function of Fermi level for
	vacancy defects in PT in four different growth conditions which were illustrated
	with A, B, C and D in figure~\ref{fig-smp2}. The zero of Fermi level corresponds to the top of
	the valence band. The slope of these lines indicates the charge state. (a), (b),
	(c) and (d) correspond to A, B, C and D, four different growth conditions,
	respectively.}
	\label{fig-smp3}
\end{figure}

The dependence of the formation energies on ${\varepsilon _\text{F}}$ for native
vacancy defects was shown in figure~\ref{fig-smp3} in four different growth conditions
which were illustrated with A, B, C and D in figure~\ref{fig-smp2}. These results were
obtained from the first-principles calculations according to equation~(\ref{eq:1}) and the
slope of the line represents the charge state of the native vacancy defect. There
is a little difference between figures~\ref{fig-smp3}~(a)~to~(d) which can be attributed to
the chemical potentials of Pb, Ti and O in four growth conditions. It could be
seen that the vacancy formation energy for Pb is generally lower than those for
others at all growth conditions as shown in figure~\ref{fig-smp3}. It could be concluded that
V$_{\text{Pb}}$'s, 
 usually caused by Pb volatilization, are most likely to be formed
during the growth process of PT. Our results are consistent with the conclusion
drawn in the experiment \cite{20}.

\subsection{Analysis of Mn defects}

\subsubsection{Energies}
\begin{figure}[!t]
	\centering
	\includegraphics[width=0.7\linewidth]{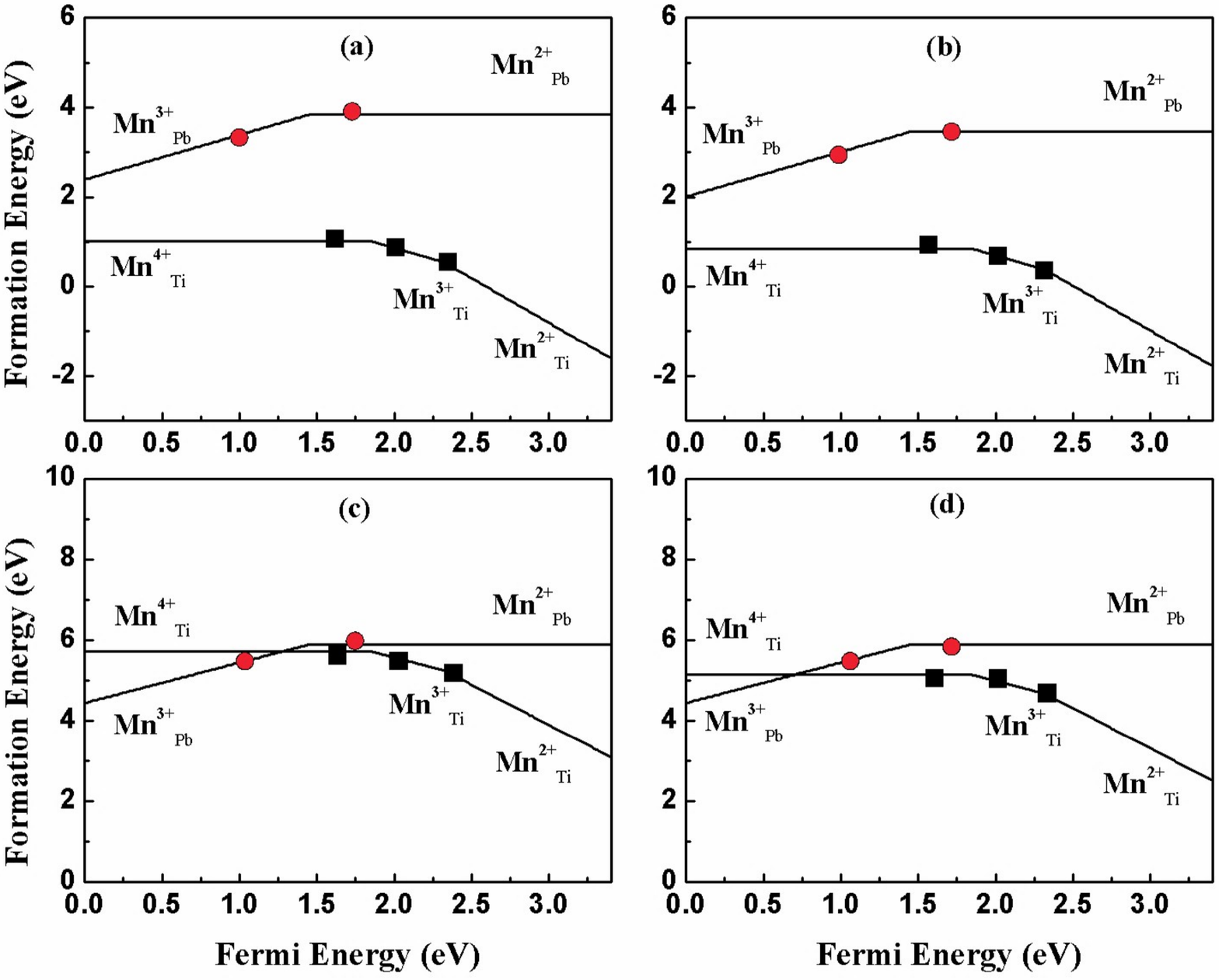}
	\caption{(Colour online) Formation energies as a function of Fermi level for
		Mn in different configurations (Pb-substitutional, Ti-substitutional) in the host
		supercell without V$_{{\rm{Pb}}}$ in different growth conditions. Kinks in the curves
		indicate transitions between different charge states. Red solid circles and black
		solid squares represent the defect formation energies of Mn$_{{\rm{Pb}}}$ and Mn$_{{\rm{Ti}}}$
		at corresponding Fermi levels in different charge states respectively. (a), (b),
		(c) and (d) correspond to A, B, C and D, four different growth conditions in figure~\ref{fig-smp2},
		respectively.}
	\label{fig-smp4}
\end{figure}

Since V$_{{\rm{Pb}}}$ is the most probable vacancy defect during the growth process of
PT, usually it is possible that a few V$_{\text{Pb}}$'s still exist in PT though the excess
Pb is added during preparation process of PT. Thus, Mn defects in PT should be
studied based on the Pb vacancy defects. In order to study the effect of V$_{\rm{Pb}}$
on Mn doping in the PT crystal, one Mn impurity is added into Pb site and Ti site
in the two kinds of ${\rm{3}} \times {\rm{3}} \times {\rm{3}}$ host supercells,
respectively. The perfect supercells were marked with model \#1 and those with
one V$_{\rm{Pb}}$ were marked with model \#2 as shown in figures~\ref{fig-smp1}~(e)
to (h).

In model \#1, V$_{{\rm{Pb}}}$ is not considered in Mn doped PT. The formation
energies of Mn defects in different valence states and sites were investigated
and discussed in this case firstly. Figure~\ref{fig-smp4} shows the formation energy as a
function of ${\varepsilon _\text{F}}$ for Mn in different configurations with four different growth
conditions. The line segments of a different slope were displayed corresponding
to different charge states and they connect each other at the corresponding
transition level for the same doping defect. The corresponding transition levels
are listed in table~\ref{tbl-smp3}. The formation energies increase with ${\varepsilon _\text{F}}$ 
and decreasing of Mn ion's valence for Mn$_{\rm{Pb}}$, while
those are contrary for Mn$_{\rm{Ti}}$. Furthermore, the neutral charge states have
obviously higher energies than other charge states for both Mn$_{\rm{Pb}}$ and
Mn$_{\rm{Ti}}$. It indicated that Mn ion doping into Pb site at a higher valence is more
stable, while it is more stable for Mn$_{\rm{Ti}}$ at a lower valence, only considering
the defect formation energies. It is also noted that the formation energies of
Mn$_{\rm{Ti}}$ are always much lower than those of Mn$_{\rm{Pb}}$ in figures~\ref{fig-smp4}~(a)
and~\ref{fig-smp4}~(b) corresponding to the oxygen-rich growth, while they increase and get
close to each other in figures~\ref{fig-smp4}~(c) and~\ref{fig-smp4}~(d) corresponding to the oxygen-poor
growth, which results from the varieties of the chemical potentials of Pb, Ti and
O in four growth conditions illustrated in figure~\ref{fig-smp2}. The formation energies
of  ${\rm{Mn}}_{{\rm{Ti}}}^{{\rm{2}} + }$ still become lower than those of Mn$_{\rm{Pb}}$ in this case.

\begin{table}[!t]
	\caption{Transition levels ${\varepsilon _{{q_1} - {q_2}}}$ (eV) of different charge states for
		Mn defects which are the kinks on the curves in figure~\ref{fig-smp4} (Mn doped PT without
		V$_{{\rm{Pb}}}$) and figure~\ref{fig-smp5} (Mn doped PT with V$_{{\rm{Pb}}}$).}
	\label{tbl-smp3} 
\begin{center}
	\vspace{3mm} \noindent
\begin{tabular}{p{15pt}p{52pt}p{52pt}p{52pt}|p{52pt}p{52pt}p{52pt}}
	\hline
	\parbox{15pt}{\centering } & \multicolumn{3}{c|}{\parbox{174pt}{\centering 
			{\small Mn doped PT without V$_{{\rm{Pb}}}$}
	}} & \multicolumn{3}{|c}{\parbox{174pt}{\centering 
			{\small Mn doped PT with V$_{{\rm{Pb}}}$}
	}} \\
	\parbox{15pt}{\centering } & \parbox{52pt}{\centering 
		{\small ${\rm{Mn}}_{{\rm{Pb}}}^{{\rm{3}}+}\to{\rm{Mn}}_{{\rm{Pb}}}^{{\rm{2}}+}$}
	} & \parbox{52pt}{\centering 
		{\small ${\rm{Mn}}_{{\rm{Ti}}}^{{\rm{4}}+}\to{\rm{Mn}}_{{\rm{Ti}}}^{{\rm{3}}+}$}
	} & \parbox{52pt}{\centering 
		{\small ${\rm{Mn}}_{{\rm{Ti}}}^{{\rm{3}}+}\to{\rm{Mn}}_{{\rm{Ti}}}^{{\rm{2}}+}$}
	} & \parbox{52pt}{\centering 
		{\small ${\rm{Mn}}_{{\rm{Pb}}}^{{\rm{3}}+}\to{\rm{Mn}}_{{\rm{Pb}}}^{{\rm{2}}+}$}
	} & \parbox{52pt}{\centering 
		{\small ${\rm{Mn}}_{{\rm{Ti}}}^{{\rm{4}}+}\to{\rm{Mn}}_{{\rm{Ti}}}^{{\rm{3}}+}$}
	} & \parbox{52pt}{\centering 
		{\small ${\rm{Mn}}_{{\rm{Ti}}}^{{\rm{3}}+}\to{\rm{Mn}}_{{\rm{Ti}}}^{{\rm{2}}+}$}
	} \\
	\hline
	\parbox{15pt}{\centering 
		{\small ${\varepsilon _{{q_1} - {q_2}}}$}
	} & \parbox{52pt}{\centering 
		{\small 1.46}
	} & \parbox{52pt}{\centering 
		{\small 1.85}
	} & \parbox{52pt}{\centering 
		{\small 2.32}
	} & \parbox{52pt}{\centering 
		{\small 1.46}
	} & \parbox{52pt}{\centering 
		{\small 0.98}
	} & \parbox{52pt}{\centering 
		{\small 1.05}
	} \\
	\hline
\end{tabular}
\end{center}
\end{table}

\begin{figure}[!t]
	\centering
	\includegraphics[width=0.7\linewidth]{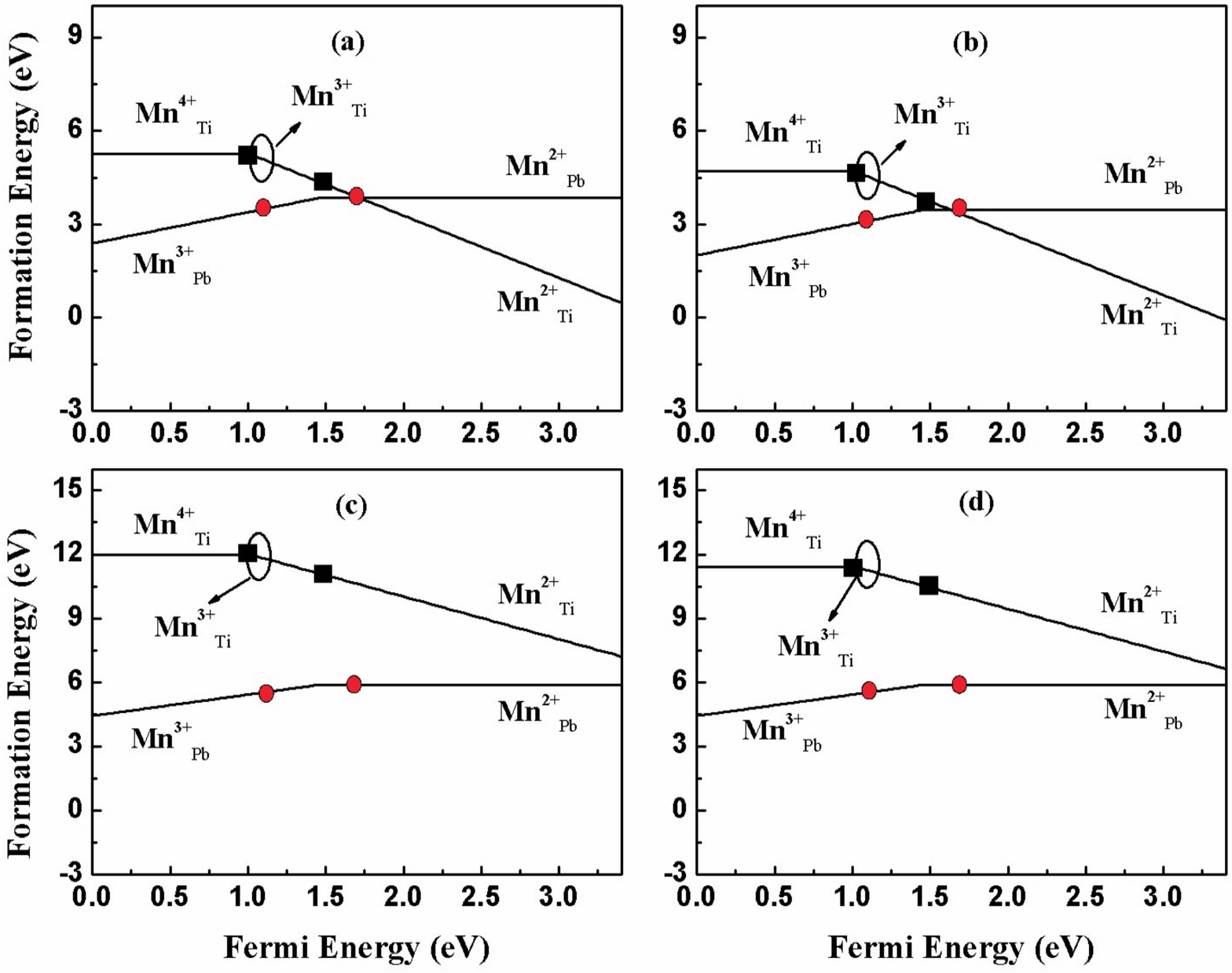}
	\caption{(Colour online) Formation energies vs Fermi energy for the defects of
		Mn$_{{\rm{Pb}}}$ and Mn$_{{\rm{Ti}}}$ at each relevant charge state in the host supercell with
		V$_{{\rm{Pb}}}$ in different growth conditions. Red solid circles and black solid squares
		have the same meaning as in figure~\ref{fig-smp4}. (a), (b),
		(c) and (d) correspond to A, B, C and D, four different growth conditions in figure~\ref{fig-smp2},
		respectively.}
	\label{fig-smp5}
\end{figure}

In another case (as described in model \#2), V$_{{\rm{Pb}}}$ is considered in the study
of Mn doped PT. When a Mn ion enters the A site, it could compensate V$_{{\rm{Pb}}}$, while if it
enters the B site, the V$_{{\rm{Pb}}}$ will be left. The formation energies for Mn in different
configurations are analyzed as a function of Fermi energy, and are shown in
figure~\ref{fig-smp5}. Comparing the formation energies of model \#1 (not containing
V$_{{\rm{Pb}}}$) in figure~\ref{fig-smp4}, it can be found that those of Mn$_{{\rm{Ti}}}$ have an increase
corresponding to the same growth conditions which results in the formation
energies of Mn$_{{\rm{Pb}}}$ that are obviously lower than those of Mn$_{{\rm{Ti}}}$ in figures~\ref{fig-smp5}~(c)
and~\ref{fig-smp5}~(d), while they get close to each other in figures~\ref{fig-smp5}~(a) and~\ref{fig-smp5}~(b).

The defect formation energies were calculated at corresponding Fermi levels for
Mn in different configurations, and listed in table~\ref{tbl-smp4} with Fermi levels and the values in~\cite{32} too. Their positions are marked with red solid circles for
Mn$_{{\rm{Pb}}}$ and with black solid squares for Mn$_{{\rm{Ti}}}$, respectively, in figures~\ref{fig-smp4} and~\ref{fig-smp5},
except that for Mn$_{{\rm{Ti}}}^{3+}$ due to too short line segment in figure~\ref{fig-smp5},
which agree with the line segment of their charge state in the diagrams.
Since the defect formation energies have a close relationship with the chemical
potential selecting, the values in this paper are slightly different from those
in the~\cite{32}. The stable charge state is the one which has the lowest formation
energy for a given Fermi level \cite{29}. The neutral charge states have obviously
higher energies than other charge states, and the formation energies of Mn$_{{\rm{Ti}}}^{2 + }$ are
the lowest in model \#1 (not containing V$_{{\rm{Pb}}}$), while
those of Mn$_{{\rm{Pb}}}^{3 + }$ are the lowest in model \#2 (containing V$_{{\rm{Pb}}}$).
Thus, considering minimization of the defect formation energy, it could be concluded
that (1) the neutral Mn defects are more unstable than those of other charge
states \cite{33}; (2) Mn ions prefer the B sites in +2 valence state in PT without
V$_{{\rm{Pb}}}$, while Mn ions prefer A sites in +3 valence state in PT with V$_{{\rm{Pb}}}$,
which shows that A sites are more favored by Mn ions when Pb is deficient.

\begin{table}[!t]
	\caption{Formation energies $E$~(eV) of different
		defects at corresponding Fermi levels ${\varepsilon _\text{F}}$~(eV) which have been
		calibrated with the top of the valence band for Mn in different configurations.
		$E_f^{\rm{A}}$, $E_f^{\rm{B}}$, $E_f^{\rm{C}}$ and $E_f^{\rm{D}}$ represent defect formation energies in A, B,
		C and D, four different growth conditions in figure~\ref{fig-smp2}, $E_f^{\rm{1}}$, 
		$E_f^{\rm{2}}$ and $E_f^{\rm{3}}$ represent those in
		three different growth conditions in~\cite{32}.}
	\label{tbl-smp4} 
	\begin{center}
		\vspace{3pt} \noindent
		\begin{tabular}{p{68pt}p{20pt}p{23pt}p{23pt}p{28pt}p{29pt}p{23pt}p{27pt}p{23pt}}
			\hline
			\parbox{68pt}{\centering 
				{\small Defect}
			} & \parbox{20pt}{\centering {\small ${\varepsilon _\text{F}}$}} & \parbox{23pt}{\centering {\small $E_f^{\rm{A}}$}} & \parbox{23pt}{\centering {\small$E_f^{\rm{B}}$}} & \parbox{28pt}{\centering {\small $E_f^{\rm{C}}$}} & \parbox{29pt}{\centering {\small $E_f^{\rm{D}}$}} & \parbox{23pt}{\centering {\small $E_f^{\rm{1}}$}} & \parbox{27pt}{\centering {\small $E_f^{\rm{2}}$}} & \parbox{23pt}{\centering {\small $E_f^{\rm{3}}$}} \\
			\hline
			\parbox{68pt}{\centering  {\small ${\rm{Mn}}_{{\rm{Pb}}}^{{\rm{2+}}}$}
			}& \parbox{20pt}{\centering 
				{\small 1.704}
			} & \parbox{23pt}{\centering 
				{\small 3.850}
			} & \parbox{23pt}{\centering 
				{\small 3.470}
			} & \parbox{28pt}{\centering 
				{\small 5.900}
			} & \parbox{29pt}{\centering 
				{\small 5.900}
			} & \parbox{23pt}{\centering 
				{\small $ \sim $2.3}
			} & \parbox{27pt}{\centering 
				{\small $ \sim $4.6}
			} & \parbox{23pt}{\centering 
				{\small $ \sim $4.6}
			} \\\\
			\parbox{68pt}{\centering {\small ${\rm{Mn}}_{{\rm{Pb}}}^{{\rm{3+}}}$}
			} & \parbox{20pt}{\centering 
				{\small 1.095}
			} & \parbox{23pt}{\centering 
				{\small 3.485}
			} & \parbox{23pt}{\centering 
				{\small 3.105}
			} & \parbox{28pt}{\centering 
				{\small 5.535}
			} & \parbox{29pt}{\centering 
				{\small 5.535}
			} & \parbox{23pt}{\centering 
				{\small --}
			} & \parbox{27pt}{\centering 
				{\small --}
			} & \parbox{23pt}{\centering 
				{\small --}
			} \\\\
			\parbox{68pt}{\centering {\small ${\rm{Mn}}_{{\rm{Ti}}}^{{\rm{2+}}}$}
			} & \parbox{20pt}{\centering 
				{\small 2.298}
			} & \parbox{23pt}{\centering 
				{\small 0.594}
			} & \parbox{23pt}{\centering 
				{\small 0.424}
			} & \parbox{28pt}{\centering 
				{\small 5.294}
			} & \parbox{29pt}{\centering 
				{\small 4.714}
			} & \parbox{23pt}{\centering 
				{\small --}
			} & \parbox{27pt}{\centering 
				{\small --}
			} & \parbox{23pt}{\centering 
				{\small --}
			} \\\\
			\parbox{68pt}{\centering {\small ${\rm{Mn}}_{{\rm{Ti}}}^{{\rm{3+}}}$}
			} & \parbox{20pt}{\centering 
				{\small 1.996}
			} & \parbox{23pt}{\centering 
				{\small 0.874}
			} & \parbox{23pt}{\centering 
				{\small 0.704}
			} & \parbox{28pt}{\centering 
				{\small 5.574}
			} & \parbox{29pt}{\centering 
				{\small 4.994}
			} & \parbox{23pt}{\centering 
				{\small --}
			} & \parbox{27pt}{\centering 
				{\small --}
			} & \parbox{23pt}{\centering 
				{\small --}
			} \\\\
			\parbox{68pt}{\centering {\small ${\rm{Mn}}_{{\rm{Ti}}}^{{\rm{4+}}}$}
			} & \parbox{20pt}{\centering 
				{\small 1.594}
			} & \parbox{23pt}{\centering 
				{\small 1.020}
			} & \parbox{23pt}{\centering 
				{\small 0.850}
			} & \parbox{28pt}{\centering 
				{\small 5.720}
			} & \parbox{29pt}{\centering 
				{\small 5.140}
			} & \parbox{23pt}{\centering 
				{\small $ \sim $0.5}
			} & \parbox{27pt}{\centering 
				{\small $ \sim -1.3$}
			} & \parbox{23pt}{\centering 
				{\small $ \sim $0.1}
			} \\\\
			\parbox{68pt}{\centering 
				{\small ${\rm{Mn}}_{{\rm{Ti}}}^{{\rm{2+}}}$ with V$_{{\rm{Pb}}}$}
			} & \parbox{20pt}{\centering 
				{\small 1.472}
			} & \parbox{23pt}{\centering 
				{\small 4.336}
			} & \parbox{23pt}{\centering 
				{\small 3.786}
			} & \parbox{28pt}{\centering 
				{\small 11.086}
			} & \parbox{29pt}{\centering 
				{\small 10.506}
			} & \parbox{23pt}{\centering 
				{\small --}
			} & \parbox{27pt}{\centering 
				{\small --}
			} & \parbox{23pt}{\centering 
				{\small --}
			} \\\\
			\parbox{68pt}{\centering 
				{\small ${\rm{Mn}}_{{\rm{Ti}}}^{{\rm{3+}}}$ with V$_{{\rm{Pb}}}$}
			} & \parbox{20pt}{\centering 
				{\small 1.206}
			} & \parbox{23pt}{\centering 
				{\small 5.024}
			} & \parbox{23pt}{\centering 
				{\small 4.474}
			} & \parbox{28pt}{\centering 
				{\small 11.774}
			} & \parbox{29pt}{\centering 
				{\small 11.194}
			} & \parbox{23pt}{\centering 
				{\small --}
			} & \parbox{27pt}{\centering 
				{\small --}
			} & \parbox{23pt}{\centering 
				{\small --}
			} \\\\
			\parbox{68pt}{\centering 
				{\small  ${\rm{Mn}}_{{\rm{Ti}}}^{{\rm{4+}}}$ with V$_{{\rm{Pb}}}$}
			} & \parbox{20pt}{\centering 
				{\small 1.009}
			} & \parbox{23pt}{\centering 
				{\small 5.250}
			} & \parbox{23pt}{\centering 
				{\small 4.700}
			} & \parbox{28pt}{\centering 
				{\small 12.00}
			} & \parbox{29pt}{\centering 
				{\small 11.420}
			} & \parbox{23pt}{\centering 
				{\small --}
			} & \parbox{27pt}{\centering 
				{\small --}
			} & \parbox{23pt}{\centering 
				{\small --}
			} \\
			\hline
		\end{tabular}
	\end{center}
\end{table}

Considering the change of defect configurations in PT with the concentration of
Mn doping, based on the conclusion from model \#2 (containing V$_{{\rm{Pb}}}$), a
few Pb deficiencies caused by Pb volatilization in PT could be relieved due
to the compensation of Mn ions entering the Pb sites in +3 valence state, which
results in a decrease of the trapping electronic charge and makes the domain wall
motion easier, thus improving ferroelectric and piezoelectric properties of PT~\cite{14,16}. Then, at a further increase of Mn doping concentration, there will be
no V$_{\text{Pb}}$'s needing Mn ions to fill, based on the conclusion from model \#1
without V$_{{\rm{Pb}}}$, more Mn ions in +2 valence state are more likely to replace
Ti$^{4+}$ to increase oxygen vacancy concentration in PT which results in a
decrease of the relative permittivity and loss tangent usually~\cite{14,15,16,17,18}. The
measurement of Electron Spin Resonance also shows that Mn ions coexist mainly in
the way of~Mn$^{2+}$ and Mn$^{3+}$ in PZT ceramics~\cite{14}.

\subsubsection{Structure}
The volumes of $3 \times 3 \times 3$ PT supercells with different defects are
listed with their change rate compared to that of pure PT in table~\ref{tbl-smp5}. The
supercell has a larger volume change for Mn doping at A site than that for Mn
doping at B site at the same valence state, which was also observed in the
experimental result of~0.5\% Mn doped PZT~\cite{16}. Additionally, it becomes larger
and larger with Mn ion's valence. The ionic radii of Pb$^{2+ }$ and Ti$^{4+}$ are
1.32~\AA{}, 0.64~\AA{}, while those of Mn$^{2+}$, Mn$^{3+}$ and Mn$^{4+}$ are
0.91~\AA{}, 0.65~\AA{} and~0.53~\AA{}, respectively \cite{34}. Consequently, Mn
substituting for Pb could produce a large lattice distortion while its
substitution for Ti$^{4+}$ causes a small lattice distortion due to a similar ionic
radius between Mn and Ti ions. Greater lattice deformation would lead to
the structural change or unstable state of PT eventually by more Mn entering the A site
which is not actually found in the experiment. Therefore, it is inferred that Mn
ion may enter the A site only at a low concentration while it mainly enters the B
site at a high concentration.

\begin{table}[!t]
	\caption{Calculated volumes of  PT $3 \times 3 \times
		3$ supercells with different defects and their corresponding change rates compared
		to that of perfect PT.}
		\vspace{3mm}
	\label{tbl-smp5} 
	\begin{center}
		\vspace{2pt} 
		\begin{tabular}{p{88pt}p{47pt}p{47pt}}
			\hline
			\parbox{88pt}{\centering 
				{\small Configuration}
			} & \parbox{47pt}{\centering 
				{\small Volume (\AA{})}
			} & \parbox{47pt}{\centering 
				{\small Change rate}
			} \\
			\hline
			\parbox{88pt}{\centering 
				{\small PT}
			} & \parbox{47pt}{\centering 
				{\small 1689.41}
			} & \parbox{47pt}{\centering 
				{\small --}
			} \\\\
			\parbox{88pt}{\centering 
				{\small PT with ${\rm{Mn}}_{{\rm{Pb}}}^{{\rm{2 + }}}$ }
			} & \parbox{47pt}{\centering 
				{\small 1676.33}
			} & \parbox{47pt}{\centering 
				{\small $-0.77$\%}
			} \\\\
			\parbox{88pt}{\centering 
				{\small PT with ${\rm{Mn}}_{{\rm{Pb}}}^{{\rm{3 + }}}$ }
			} & \parbox{47pt}{\centering 
				{\small 1668.77}
			} & \parbox{47pt}{\centering 
				{\small $-1.22$\%}
			} \\\\
			\parbox{88pt}{\centering 
				{\small PT with ${\rm{Mn}}_{{\rm{Ti}}}^{{\rm{2 + }}}$}
			} & \parbox{47pt}{\centering 
				{\small 1694.70}
			} & \parbox{47pt}{\centering 
				{\small 0.31\%}
			} \\\\
			\parbox{88pt}{\centering 
				{\small PT with ${\rm{Mn}}_{{\rm{Ti}}}^{{\rm{3 + }}}$ }
			} & \parbox{47pt}{\centering 
				{\small 1684.88}
			} & \parbox{47pt}{\centering 
				{\small $-0.27$\%}
			} \\\\
			\parbox{88pt}{\centering 
				{\small PT with ${\rm{Mn}}_{{\rm{Ti}}}^{{\rm{4 + }}}$ }
			} & \parbox{47pt}{\centering 
				{\small 1676.41}
			} & \parbox{47pt}{\centering 
				{\small $-0.77$\%}
			} \\\\
			\parbox{88pt}{\centering 
				{\small PT with ${\rm{Mn}}_{{\rm{Ti}}}^{{\rm{2 + }}}$ and V$_{{\rm{Pb}}}$}
			} & \parbox{47pt}{\centering 
				{\small 1688.36}
			} & \parbox{47pt}{\centering 
				{\small $-0.06$\%}
			} \\\\
			\parbox{88pt}{\centering 
				{\small PT with ${\rm{Mn}}_{{\rm{Ti}}}^{{\rm{3 + }}}$ and V$_{{\rm{Pb}}}$}
			} & \parbox{47pt}{\centering 
				{\small 1680.00}
			} & \parbox{47pt}{\centering 
				{\small $-0.56$\%}
			} \\\\
			\parbox{88pt}{\centering 
				{\small PT with ${\rm{Mn}}_{{\rm{Ti}}}^{{\rm{4 + }}}$ and V$_{{\rm{Pb}}}$}
			} & \parbox{47pt}{\centering 
				{\small 1671.67}
			} & \parbox{47pt}{\centering 
				{\small $-1.05$\%}
			} \\
			\hline
		\end{tabular}
		\vspace{2pt}
		
	\end{center}
\end{table}

\subsubsection{Bader charge}
\begin{table}[!t]
	\caption{Average Bader charges ($e$) of Pb, Ti and O of PT for
		different Mn doping, respectively, and their changes $\Delta q$ compared with those of undoped PT, are given in parentheses.}
	\label{tbl-smp6}
	\vspace{3mm} 
	\begin{center}
		\vspace{3pt}
		\begin{tabular}{p{33pt}p{44pt}p{54pt}p{54pt}p{54pt}p{54pt}}
			\hline
			\parbox{33pt}{\centering \strut
				{\small Atoms}
			} & \parbox{44pt}{\centering 
				{\small undoping}
			} & \parbox{54pt}{\centering 
				{\small ${\rm{Mn}}_{{\rm{Pb}}}^{{\rm{2 + }}}$ ($\Delta q$)}
			} & \parbox{54pt}{\centering 
				{\small ${\rm{Mn}}_{{\rm{Pb}}}^{{\rm{3 + }}}$ ($\Delta q$)}
			} & \parbox{54pt}{\centering 
				{\small ${\rm{Mn}}_{{\rm{Ti}}}^{{\rm{2 + }}}$ ($\Delta q$)}
			} & \parbox{54pt}{\centering 
				{\small ${\rm{Mn}}_{{\rm{Ti}}}^{{\rm{3 + }}}$ ($\Delta q$)}
			} \\ 
			\hline \\
			\parbox{33pt}{\centering 
				{\small Pb}
			} & \parbox{44pt}{\centering 
				{\small 1.35644}
			} & \parbox{54pt}{\centering 
				{\small 1.340324}
				{\small ($-0.016116$)}
			} & \parbox{54pt}{\centering 
				{\small 1.34888}
				{\small ($-0.00756$)}
			} & \parbox{54pt}{\centering 
				{\small 1.334463}
				{\small ($-0.021977$)}
			} & \parbox{54pt}{\centering 
				{\small 1.341407}
				{\small ($-0.015033$)}
			} \\\\
			\parbox{33pt}{\centering 
				{\small Ti}
			} & \parbox{44pt}{\centering 
				{\small 1.846502}
			} & \parbox{54pt}{\centering 
				{\small 1.856993}
				{\small (0.010491)}
			} & \parbox{54pt}{\centering 
				{\small 1.859499}
				{\small (0.012997)}
			} & \parbox{54pt}{\centering 
				{\small 1.840132}
				{\small ($-0.00637$)}
			} & \parbox{54pt}{\centering 
				{\small 1.848042}
				{\small (0.00154)}
			} \\\\
			\parbox{33pt}{\centering 
				{\small O}
			} & \parbox{44pt}{\centering 
				{\small $-1.067647$}
			} & \parbox{54pt}{\centering 
				{\small $-1.06498$}
				{\small (0.002667)}
			} & \parbox{54pt}{\centering 
				{\small $-1.057343$}
				{\small (0.010304)}
			} & \parbox{54pt}{\centering 
				{\small $-1.078887$}
				{\small ($-0.01124$)}
			} & \parbox{54pt}{\centering 
				{\small $-1.072041$}
				{\small ($-0.004394$)}
			} \\
			\hline
		\end{tabular}
	\end{center}
\end{table}

\begin{table}[!t]
	\caption{The Bader charges ($e$) of Mn ions in different
		valence states at A and B sites of PT. ${N_{1}}$, ${N_{2}}$ and $\Delta N$
		represent the electronic numbers of Mn
		atom, the Bader charges of Mn ions and ${N_{2}} - {N_{1}}$.}
	\label{tbl-smp7} 
		\vspace{3mm} 
\begin{center}
	\vspace{3pt} \noindent
	\begin{tabular}{p{68pt}p{80pt}p{67pt}p{103pt}}
		\hline
		\parbox{68pt}{\centering 
			{\small	Defect} 
		} & \parbox{80pt}{\centering {\small ${N_{\rm{1}}}$}} & \parbox{67pt}{\centering {\small ${N_{\rm{2}}}$}} & \parbox{103pt}{\centering {\small $\Delta N$}} \\
		\hline
		\parbox{68pt}{\centering {\small ${\rm{Mn}}_{{\rm{Pb}}}^{{\rm{2 + }}}$ }} & \parbox{80pt}{\centering 
			{\small 7.00}
		} & \parbox{67pt}{\centering 
			{\small 5.723836}
		} & \parbox{103pt}{\centering 
			{\small 1.276164}
		} \\\\
		\parbox{68pt}{\centering {\small ${\rm{Mn}}_{{\rm{Pb}}}^{{\rm{3 + }}}$}} & \parbox{80pt}{\centering 
			{\small 7.00}
		} & \parbox{67pt}{\centering 
			{\small 5.632596}
		} & \parbox{103pt}{\centering 
			{\small 1.367404}
		} \\\\
		\parbox{68pt}{\centering {\small ${\rm{Mn}}_{{\rm{Ti}}}^{{\rm{2 + }}}$}} & \parbox{80pt}{\centering 
			{\small 7.00}
		} & \parbox{67pt}{\centering 
			{\small 5.484108}
		} & \parbox{103pt}{\centering 
			{\small 1.515892}
		} \\\\
		\parbox{68pt}{\centering  {\small${\rm{Mn}}_{{\rm{Ti}}}^{{\rm{3 + }}}$}} & \parbox{80pt}{\centering 
			{\small 7.00}
		} & \parbox{67pt}{\centering 
			{\small 5.431728}
		} & \parbox{103pt}{\centering 
			{\small 1.568272}
		} \\\\
		\parbox{68pt}{\centering  {\small ${\rm{Mn}}_{{\rm{Ti}}}^{{\rm{4 + }}}$}} & \parbox{80pt}{\centering 
			{\small 7.00}
		} & \parbox{67pt}{\centering 
			{\small 5.374644}
		} & \parbox{103pt}{\centering 
			{\small 1.625356}
		} \\
		\hline
	\end{tabular}
\end{center}
\end{table}
Average Bader charges of Pb, Ti and O of PT with and without Mn doping at
different sites are listed in table~\ref{tbl-smp6}, respectively. One Ti ion and one Pb ion
correspond to three O ions according to stoichiometric ratio in PT and the
increase of positive charge of Ti and Pb ions always corresponds to the increase
of negative charge of O ions which could be verified by average Bader charges of
Pb, Ti and O ($1.35644+1.846502 \approx 3\times1.067647$). The charge number of O ion
is reduced for Mn entering the A-site compared with undoped PT which indicates
that the electronegativity of O becomes weaker. The electronegativity of O could
affect the binding force between ions. Thus, when Mn enters the A site, the binding
force between ions becomes weaker, which makes the domain movement easier in PT
to improve the electromechanical performance of PT, while Mn entering the B site is the opposite.

The bonding mode of Mn ion entering the different position of PT directly affects
the charge distribution of PT which results in the change of its performance. The
Bader charges of Mn ions at different lattice sites are listed in table~\ref{tbl-smp7},
respectively. Mn ions lost more electrons entering the B site than entering the A site, which
shows that the ionicity of the chemical bond is stronger when Mn ion enters the B
site. In addition, the ionicity of Mn increases with its valence.

\section{Conclusion}
Mn defects in PT have been studied by the first-principles based on the native
vacancy defects. The defects of Mn$_{{\rm{Pb}}}$ and Mn$_{{\rm{Ti}}}$ were analyzed in two
different lattice models with and without V$_{\text{Pb}}$s which are the most probable
vacancy defects to occur during the growth process of PT. It is found that
whether Mn ion enters A or B site of PT is related to its doping concentration.
PT prefers to incorporate Mn on Pb sites rather than Ti sites with the Pb
deficiencies at a low Mn concentration, while a B-site Mn doping is preferable at a
high concentration. Mn ion entering the A site with $+3$ valence could result in the
lattice distortion of PT becoming larger and the electronegativity of O becoming
weaker, which makes the domain movement easier in PT to improve the performance of
PT.

\section*{Acknowledgements}
The work was supported by the National Natural Science Foundation of China (No.
11705136), the Special Foundation of Shaanxi Educational Commission (No.
17JK0435), the Natural Science Foundations of Shaanxi Province of China (No.
2018JQ1048 and No. 2020JQ-656).

\newpage
\ukrainianpart

\title{Центри легування і валентність іонів Mn у PbTiO$_{3}$ на основі природних дефектів вакансій
}
\author{Г. Сін, К. Панг, Д.Л. Гао, Л. Лі, П. Жанг
}
\address{
	Науковий коледж університету архітектури і технологій м. Ксян, Ксян, 710055, КНР
	}

\makeukrtitle 

\begin{abstract}
На основі першопринципного моделювання вивчалися місце легування і валентність іонів Mn у сполуках PbTiO$_{3}$ (РТ) з природними дефектами вакансій. Спершу були вивчені природні дефекти вакансій Pb, O та Ti у PT, і було виявлено, що вакансія Pb є найкращою. При рості РТ, легованої Mn, переважним має бути заміщення іонами Mn з валентністю +3 іонів Pb в А-центрах, коли концентрація Pb є недостатньою в умовах рівноваги. Це обумовлено виключно мінімізацією енергії утворення і може призвести до більшої дисторсії гратки РТ. Крім того, коли Mn потрапляє на місце Pb, електронегативність O стає слабшою, що полегшує рух доменів в PT і покращує характеристики цієї сполуки, тоді як заміщення іонами Mn іонів Ti в B-центрі має протилежний ефект.
	\keywords легування Mn, сполуки PT, природні дефекти, енергія утворення дефектів, заряд Бейдера
\end{abstract}


\begin{thebibliography}{34}
\bibitem{1} Niu~X.~S., Jia~W., Qian~S., Zhu~J., Zhang~J., Hou~X.~J., Mu~J.~L., Geng~W.~P., Cho~J.~D., He~J., Chou~X.~J., ACS Sustainable Chem. Eng., 2019, \textbf{7}, 979--985,		\doi{10.1021/acssuschemeng.8b04627}.
\bibitem{2} Engholm~M., Bouzari~H., Christiansen~T.~L., Beers~C., Bagge~J.~P., Moesner L.~N., Diederichsen~S.~E., Stuart~M.~B., Jensen~J.~A., Thomsen~E.~V., Sens. Actuators, A, 2018, \textbf{273}, 121--133, \doi{10.1016/j.sna.2018.02.031}.
\bibitem{3} Kudela P., Radzienski M., Ostachowicz W., Yang Z.~B., Mech. Syst. Sig. Process., 2018, \textbf{108}, 21--32, \\ \doi{10.1016/j.ymssp.2018.02.008}.		
\bibitem{4} Wang H., Chen Z.~F., Xie H.~K., Sens. Actuators, A, 2020, \textbf{3091}, 112018, \doi{10.1016/j.sna.2020.112018}.		
\bibitem{5} Li J.~F., Zhu Z.~X., Lai F.~P., J. Phys. Chem. C, 2010, \textbf{114}, 17796--17801, \doi{10.1021/jp106384e}.		
\bibitem{6} Wongdamnern N., Triamnak N., Ngamjarurojana A., Laosiritaworn
		Y., Ananta S., Yimnirun R., Ceram. Int., 2008, \textbf{34}, 731--734, 
		\doi{10.1016/j.ceramint.2007.09.048}.		
\bibitem{7} Tong S., Ma B.~H., Narayanan M., Liu S., Koritala R.,
		Balachandran U., Shi D., ACS Appl. Mater. Interfaces, 2013, \textbf{5}, 1474--1480,
		\doi{10.1021/am302985u}.		
\bibitem{8} Hu G.~L., Ma C. R., Wei W., Sun Z.~X., Lu L., Mi S.~B., Liu M., Ma
		B.~H., Judy W., Jia C.~L., Appl. Phys. Lett., 2016, \textbf{109}, 193904, 
		\doi{10.1063/1.4967223}.		
\bibitem{9} Mahato D.~K., Molak A., Szeremeta A.~Z., Mater. Today: Proc.,
		2017, \textbf{4}, 5488--5496,\\ \doi{10.1016/j.matpr.2017.06.004}.		
\bibitem{10} Wang Z., Ren W., Ren J.~B., Wu X., Shi P., Chen X., Yao X., Ferroelectrics, 2009, \textbf{383}, 151--158,\\
		\doi{10.1080/00150190902889333}.		
\bibitem{11} Kozielski L., Adamczyk M., Erhart J., Pawe\l{}czyk M., J.
		Electroceram., 2012, \textbf{29}, 133--138, \\\doi{10.1007/s10832-012-9746-z}.		
\bibitem{12} Zhang M.~F., Wang Y., Wang K.~F., Zhu J.~S., Liu J.~M., J. Appl.
		Phys., 2009, \textbf{105}, 061639, \doi{10.1063/1.3055338}.		
\bibitem{13} Feigl L., Pippel E., Pintilie L., Alexe M., Hesse D., J. Appl.
		Phys., 2009, \textbf{105}, 126103, \doi{10.1063/1.3141733}.		
\bibitem{14} He L.~X., Li C.~E., J. Mater. Sci., 2000, \textbf{35}, 2477--2480, 
		\doi{10.1023/A:1004717702149}.		
\bibitem{15} Liu Y.~B., Xu Z., Li Z.~R., Zhuang Y.~Y., Tian Y., Hu D., Song
		K.~X., Guo H.~S., J. Alloys Compd., 2018, \textbf{742}, 958--965,
		\doi{10.1016/j.jallcom.2018.01.027}.		
\bibitem{16} Xin H., Ren W., Wu X.~Q., Shi P., J. Appl. Phys., 2013,  \textbf{114},
		027017, \doi{10.1063/1.4812226}.		
\bibitem{17} Hennings D., Pomolun H., J. Am. Ceram. Soc., 1974, \textbf{57}, 527--532,
		\doi{10.1111/j.1151-2916.1974.tb10802.x}.		
\bibitem{18} Yu Y., Wu J.~G., Zhao T.~L., Dong S.~X., Gu H.~S., Hu Y.~M., J. Alloys Compd., 2014, 
        \textbf{615}, 676--682,\\ \doi{10.1016/j.jallcom.2014.06.144}.		
\bibitem{19} Limpijumnong S., Rujirawat S., Boonchun A., Smith M.~F., Cherdhirunkorn B., 
		Appl. Phys. Lett., 2007,\\ \textbf{90}, 103113, \doi{10.1063/1.2711200}.		
\bibitem{20} Wang L., Yu J., Wang Y., Gao J., J. Mater. Sci.: Mater. Electron., 2008, \textbf{19}, 1191--1196, \\\doi{10.1007/s10854-007-9524-x}.		
\bibitem{21} Dimos D., Schwartz R.~W., Lockwood S.~J., J. Am. Ceram. Soc.,
		1994, \textbf{77}, 3000--3005, \\\doi{10.1111/j.1151-2916.1994.tb04536.x}.		
\bibitem{22} Wang J.~L., Tang G., Wu X.~S., Pu L., J. Mater. Sci., 2014, \textbf{49},
		4715--4721, \doi{10.1007/s10853-014-8171-x}.		
\bibitem{23} Kresse G., Joubert D., Phys. Rev. B, 1999, \textbf{59}, 1758,
		\doi{10.1103/PhysRevB.59.1758}.		
\bibitem{24} Kresse G., Furthm\"{u}ller J., Phys. Rev. B,
		1996, \textbf{54}, 11169, \doi{10.1103/PhysRevB.54.11169}.
\bibitem{25} Perdew J.~P., Burke K., Ernzerhof M., Phys. Rev.
        Lett., 1996, \textbf{78}, 1396, \doi{10.1103/PhysRevLett.78.1396}.				
\bibitem{26} Robertson J., Warren W.~L., Tuttle B.~A., J. Appl. Phys., 1995,
		\textbf{77}, 3975--3980, \doi{10.1063/1.358580}.		
\bibitem{27} Xiao P., Fan X.~L., Liu L.~M., Lau W.~M., Phys. Chem. Chem. Phys., 2014, 
        \textbf{16}, 24466--24472,\\ \doi{10.1039/c4cp03453h}.		
\bibitem{28} Fang Y.~Z., Kong X.~J., Wang D.~T., Cui S.~X., Liu J.~H., Chin. J. Phys.,
		2018, \textbf{56}, 1370--1377,\\ \doi{10.1016/j.cjph.2018.04.011}.		
\bibitem{29} Van de Walle C.~G., Neugebauer J., J. Appl. Phys., 2004, \textbf{95},
		3851--3879, \doi{10.1063/1.1682673}.		
\bibitem{30} Kubota T., Nakamura A., Toyoura K., Matsunaga K., Acta
		Biomater., 2014, \textbf{10}, 3716--3722,\\ \doi{10.1016/j.actbio.2014.05.007}.		
\bibitem{31} Buckeridge J., Scanlon D.~O., Walsh A., Catlow C.~R.~A., Comput.
		Phys. Commun., 2014, \textbf{185}, 330--338,\\ \doi{10.1016/j.cpc.2013.08.026}.		
\bibitem{32} Boonchun A., Smith M.~F., Cherdhirunkorn B., Limpijumnong S., J. Appl.
		Phys., 2007, \textbf{101}, 043521,\\ \doi{10.1063/1.2654120}.		
\bibitem{33} Hayashi K., Ando A., Hamaji Y., Sakabe Y., Jpn. J. Appl. Phys.
		Part 1, 1998, \textbf{37}, 5237, \doi{10.1143/JJAP.37.5237}.		
\bibitem{34} Shannon R.~D., Acta Crystallogr., Sect. A: Found. Crystallogr., 1976, \textbf{A32}, 751--767,\\
		\doi{10.1107/S0567739476001551}.
\end{thebibliography}
\end{document}